\documentclass[english,aps,preprint,epsfig]{revtex4}
\usepackage{graphicx}
\usepackage{psfig,epsfig}
\usepackage[latin1]{inputenc}
\usepackage{babel}

\begin{document}
%\twocolumn[\hsize\textwidth\columnwidth\hsize\csname@twocolumnfalse%
%\endcsname
\title{\bf  Why RKKY exchange integrals are inappropriate to  describe ferromagnetism in diluted magnetic semiconductors.}

\author{Richard ~Bouzerar$^{1}$\footnote[4]{email: richard.bouzerar@u-picardie.fr}, Georges ~Bouzerar$^{2}$\footnote[6]{email: georges.bouzerar@grenoble.cnrs.fr} and Timothy Ziman$^{3}$\footnote[5]{and CNRS, email: ziman@ill.fr} 
}

\affiliation{ 
$^{1}$Universit\'e  de Picardie Jules Verne, 33 rue Saint-Leu, 80039 Amiens Cedex 01\\
$^{2}$Laboratoire Louis N\'eel 25 avenue des Martyrs, CNRS, B.P. 166 38042 Grenoble Cedex 09
France.\\   
$^{3}$Institut Laue Langevin B.P. 156 38042 Grenoble 
France.\\
}            
\date{\today}

\begin{abstract}
\parbox{14cm}{\rm}
\medskip
We calculate Curie temperatures and study  the stability of ferromagnetism in diluted magnetic materials,
taking as a model for the exchange between magnetic impurities a damped Ruderman-Kittel-Kasuya-Yosida (RKKY) interaction and a short range term
representing the effects of superexchange.
To properly include effects of  spin and  thermal fluctuations as well 
as  
geometric disorder, we solve the effective Heisenberg Hamiltonian 
by means of  a recently developed semi-analytical approach. This approach, ``self-consistent local Random Phase Approximation (SC-LRPA)'', is explained.
We show that previous  mean-field  treatments, which have been  widely used in the literature,
largely overestimate both the Curie temperatures and the stability of ferromagnetism as a function of    carrier density. The discrepancy when compared
to the current approach was that  effects of frustration in RKKY oscillations had been  strongly underestimated by such  simple mean-field theories.
We  argue that  the use, as is frequent, of a weakly-disordered  RKKY exchange  to model ferromagnetism in  diluted III-V  systems is inconsistent with the observation of ferromagnetism over a  wide region of itinerant carrier  densities. 
This may be puzzling when compared to 
the apparent success of calculations based on {\it ab-initio} estimates of the 
coupling; we propose a resolution to this issue
by taking  RKKY-like  interactions between 
resonant states close to the Fermi level.
\end{abstract} 

\maketitle

\section{Introduction}

 After the discovery by Ohno et al. \cite{Ohno} that the doping by a small amount of magnetic impurities in GaAs 
could lead to relatively high Curie temperature in III-V semiconductors\cite{Edmonds,Potashnik},
materials of this class have attracted
considerable interest from both experimentalists and theoreticians. 
It is now  accepted
 that the ferromagnetism is mediated by the ``itinerant carriers'' introduced after substitution of Ga, of nominal valence ${3+}$, 
by Mn$^{2+}$. 
Among theories of  the ferromagnetism in these materials,
 the simplest and most commonly  used approach  considers that the exchange between localized magnetic
 impurities is of standard RKKY type \cite{Dietl,DasSarma}. 
The exchange integrals are, to second order in perturbation theory, 
quadratic with respect to the local coupling 
of the itinerant carrier to the localized magnetic impurity spin $J_{pd}$.
\begin{eqnarray}
J_{ij}=-{J_{pd}^2 \over \pi}(\Im \ \chi({\bf R}_{ij}))
\end{eqnarray}
 where $\chi$ is the magnetic susceptibility.
The density of carriers enters in the magnetism via the dependence of the 
susceptibility but in the simplest approach
the effects of disorder (multiple scattering of the itinerant carrier due to the magnetic impurities)
are  not taken into account. 
Once the  effective magnetic Heisenberg Hamiltonian has been determined,
it has commonly been solved by treating fluctuations within mean field theory,
and disorder within a virtual crystal approximation (VCA), giving an approximation
we shall refer to as
  (MF-VCA)\cite{Dietl,DasSarma,Matsukura,Sato,Jungwirth} and is commonly referred
to loosely as Zener Mean-Field theory. It has often been stated
in the literature\cite{Dogma} that such an approach gives good account of the 
experimental situation in both III-V and II-VI doped semiconductors
and that it can be relied on for quantitative prediction.
This apparent consensus on the applicability of such a picture
is, however, seriously in doubt, as we shall discuss.

From an experimental point of view, it is  observed that the Curie temperatures are  very sensitive to the method of preparation.
 Indeed the  Curie temperatures  often vary greatly  when measured on  samples as they are grown, and after they have been annealed. In the Zener Mean-Field
Theory, the Curie temperature for fixed concentration of impurities varies with a simple power law (T$_C \propto \rho^{\frac{1}{3}})$
on the density of itinerant carriers. In reality, other aspects  are important:
such as configurational disorder, thermal fluctuations, and detailed compensating mechanisms\cite{Wang,Kirby}. 
Calculations that treat the effects of band structure realistically
from  first principles 
show that the RKKY description is inaccurate \cite{Sandraskii,DDDasSarma} in that the strength of interactions do not simply depend
on distance, but also on lattice direction.
Furthermore simple one-band model calculations of the 
exchange integrals treating the disorder by  a Coherent Potential Approximation (CPA)  has shown that the RKKY behaviour is 
in fact restricted to very small value of the ratio $J_{pd}/W$ \cite{Bouzerar1}
where $W$ is the bandwidth of the carrier band.
To higher order the oscillatory behaviour is strongly reduced. 
Calculations which, in addition, treat the effect of disorder on the itinerant
carriers within CPA, show clearly  that the exchange integrals (i) do not  oscillate and (ii) are exponentially damped with distance\cite{Bouzerar3}
as in the model calculations\cite{Bouzerar-unpublished}. 
A different approach
was taken by Brey et al\cite{Brey} who took
a {\bf k . p} description of the band structure with a {\it non-local} J$_{pd}$ coupling
treated perturbatively. In this case the non-locality of the  J$_{pd}$ coupling
lead to suppression of oscillations. Another attempt\cite{Timm}  to use
a realistic  (Slater-Koster) band structure  with non-local
couplings between p- and d-orbitals treated perturbatively,
lead, however,  to very  different
magnetic couplings which oscillated strongly. Both calculations included
spin-orbit couplings but differed in their conclusions as to its
importance. Recent work\cite{Zhou} including Monte-Carlo simulations
concluded that the effects of spin-orbit interactions on the 
Curie temperature are weak, unless the anisotropy induced is very large
which seems unlikely for III-V semiconductors. 
Fiete et al\cite{Fiete} found effects of non-collinearity from spin-orbit
coupling but they were significant  at parameters corresponding to  very low concentrations
of dilute Ga(Mn)As.

 Recent theoretical studies have shown that if the exchange integrals includes both the effects 
of disorder and realistic band structure, MF-VCA treatment of the Heisenberg model leads to larger Curie
 temperature than experimentally observed\cite{Josef}. In contrast, it was shown
that using a self-consistent local Random Phase Approximation  (SC-LRPA)
it is possible to attain
 quantitatively accurate estimates of  ferromagnetism
 in diluted magnetic semiconductors for annealed \cite{Bouzerar2} and partially
annealed samples \cite{Bouzerar4}.
This method will be  made explicit
in the next section, but the essential improvement over MF-VCA
is that 
both disorder and transverse
fluctuations  are properly treated. For the  disorder
this is 
by avoiding any effective medium approximation and, instead, using sampling
over randomly generated geometries. The tranverse fluctuations
are included 
by preserving the rotational spin symmetry in the decoupling. 
The success of this semi-analytical approach is fully supported in the 
limit of 
large spin where  
classical Monte Carlo calculations can be performed\cite{mc1,mc2} 
For finite spin we can compare only for the case
of regular lattices, but by comparison to series expansions
the Callen approach is known to give critical temperatures
accurate to a few per cent\cite{Callen} even for couplings of short range. Note that, compared to Monte-Carlo simulations, the SC-LRPA has several advantages:  it provides a direct expression for the Curie temperature and the computing cost is extremely low, allowing systematic examination
of the  space of couplings, and effects of cut-off and so forth. In addition it can include quantum fluctuations
for finite spin S for which Monte Carlo methods encounter problems of 
sign.
The disadvantage of using  ``realistic'' values of the exchange integrals 
taken from first principle calculations is that they are strongly material 
dependent, and do not provide  a simple understanding of the parameters which control the exchange integrals. Thus,
 attempts to find simplified models where the exchange integrals between magnetic impurities 
 would depend on a few  physical parameters are still  attractive and should be pursued. 

The purpose of this paper is then to return to the simple RKKY
approach whose validity has been argued on the basis of over-simplified
calculation of the thermodynamics, but now to study the 
the influence of disorder and transverse fluctuations 
using a more reliable calculation.  
Thus we shall  examine  the stability of 
ferromagnetism and Curie temperature for  exchange integrals in the effective Heisenberg model of 
RKKY type.  We will also, in the following, analyze both the effect of including 
a nearest-neighbour antiferromagnetic (AF) superexchange and damping of the exchange coupling. The damping  mimics in a simple manner the
 effect of multiple scattering of the itinerant carriers over the magnetic impurities. As we have mentioned, while this corresponds to 
the {\it average} exchange in an approximation of weak coupling
and weak disorder, it is not a  completely general form for strong coupling.

The effective Heisenberg Hamiltonian reads, 
\begin{eqnarray}
H &=& {-} \sum_{ij} {J_{ij} \bf{\Large S}_{i} \cdot \bf{\Large S}_{j}}
\end{eqnarray}                                   
where $S_i$ are quantum spins randomly distributed on a  lattice, we denote x the density of magnetic impurities. For comparison with Ga(Mn)As we take
the lattice to be face centered cubic (fcc)  corresponding to simple
substitution of Mn on Ga sites.
$J_{ij}$ is the exchange interaction coupling between two impurities located at site i and j, it reads 

\begin{eqnarray}
J_{\rm ij} &=&  J_{0} \exp({-r \over r_{0}}) \frac{({sin(2k_{F}r)-2k_{F}rcos(2k_{F}r)})} {(r/a)^{4}} + J_{ij}^{AF}
\end{eqnarray}

where $r=R_{i}-R_{j}$ is the distance between two impurities labeled i and j. The first term is the damped RKKY exchange coupling,
$r_0$ is the damping length. The second term is the direct antiferromagnetic superexchange interaction. For simplicity we restrict 
ourselves to short range SE term, $J_{ij}^{AF}=J_{AF}$ if i and j are nearest neighbours (NN) and vanishes otherwise.
We consider the diluted regime for magnetic impurities and itinerant carriers; thus we assume a spherical Fermi surface and
$k_{\rm F} = (3 \pi ^2 n_{c})^{1\over 3}$ where $n_{c}$ is the density of carriers.
Note that we shall restrict ourselves to the simplest RKKY form corresponding
to the asymptotic form of a single parabolic band of carriers. 
Our model has four parameters : $x$,$n_{c}$,$r_0$, and $J_{AF}/J_{0}$.
In the following calculations they are treated as being independent, although of course physically if we vary the 
doping $x$  to model a series of compounds of different doping, the other
parameters would be functions of $x$. A derived parameter useful
for comparison with experiment is the doping concentration {\it per}
doping impurity
 $\gamma=n_{c}/x$.  In the simplest model of doping this would be 1,
in practice it is usually much less and varies with sample history.

 As we treat the disorder
fully and thermal fluctuations accurately, we will show that previous treatments of the same Hamiltonian (Equation 2)
by MF-VCA\cite{DasSarma}  lead to wrong conclusions. 
Furthermore by understanding
this simple model,  we shall show that we can demonstrate
the importance of qualitative aspects.
In particular,
the presence or absence of oscillations will be seen to be crucial
to stabilize ferromagnetism at finite temperatures. In MF-VCA this 
was obscured by the simplified treatment.
This  may be significant in  more complicated forms of model band structures, e.g. 6 band
Kohn-Luttinger forms \cite{Dietl97,Dietl,Abolfath,Brey,Timm}. 
In addition, in different {\it ab-initio} treatments, 
for example  local-density-approximation (LDA)   or extensions such as LDA+U approaches, there are changes
in the nature of the states close to the Fermi surface\cite{DDDasSarma,Shick} which
will affect long-range  oscillations of the exchange couplings
and thus the region of stability of ferromagnetism.

One aspect we do {\it not} consider is the role of anisotropic couplings 
in spin-space coming from spin-orbit couplings. 
As we have noted above, there is some debate\cite{Timm,Brey,Zarand,Fiete} 
as to whether these make significant contributions
to the exchange.
%These have been argued to be important for the magnetic exchange in ref \cite{Timm}
%but the opposite conclusion was reached  in \cite{Brey,Zarand}.
% We argue, that in agreement with \cite{Zhou} who used Monte Carlo simulations
 We argue that for the Curie temperature
 this is not of primary importance
especially as any spin anisotropy is random in direction\cite{Fiete}.
Of
course spin-orbit effects must be considered for calculations
 of magnetic anisotropies. We will argue that for the Curie temperature,
which is the subject of the present paper, what is much more
significant is the  
bias in the sub-asymptotic form of the isotropic RKKY-like  couplings,
coming from resonant effects. This will be explained in Section IV.
The importance of such  effects, rather than anisotropies,
is supported  by the success\cite{Bouzerar2,Bouzerar4} of the  non-relativistic
{\it ab-initio } calculations  which 
include the effects of hybridization and resonance.

\section{Method}
Let us now summarize the main steps of the approach we use.
We define the following retarded Green's function $G_{ij}(t)$ for localized spins at sites  $i$ and $j$,
\begin{eqnarray}
G_{ij}(t)&=& -i\theta (t) < [S_{i}^{+}(t) ; S_{j}^{-}(0)] >
\end{eqnarray}

%Its Fourier transform in energy space is 
%\begin{eqnarray}
%G_{ij}(\omega) = \int^{+\infty}_{-\infty}{G_{ij}(t)e^{i\omega t}  } dt
%\end{eqnarray}

The important point is that we shall decouple in real space, and within a Random Phase Approximation, the equation of motion for the 
frequency-transformed Green's functions $G_{ij}(\omega)$:
\begin{eqnarray}
(\omega -h^{eff}_{i})G_{ij}(\omega) &=&2 \langle S_{i}^{z} \rangle \delta_{ij}-\left[\langle S_{i}^{z} \rangle \sum_lJ_{il}G_{lj}(\omega)\right]
\end{eqnarray}

where $\langle S_{i}^{z} \rangle$ is the local magnetization at site i. For a given configuration of impurities and temperature, $\langle S_{i}^{z} \rangle$ should be determined self-consistently at each impurity site. $h^{eff}_{i}$ is the local effective field at site i, 
\begin{eqnarray}
h^{eff}_{i} =  \sum_lJ_{il} \langle S_{l}^{z} \rangle
\end{eqnarray}

To determine $\langle S_{i}^{z} \rangle$ {\it self-consistently} we use the Callen expression: \cite{Callen},

\begin{eqnarray}
\langle S_{i}^{z} \rangle=\frac{(S-\Phi_{i})(1+\Phi_{i})^{2S+1} + (S+1+\Phi_{i}) \Phi_{i}^{ 2S+1}}
{(1+\Phi_{i})^{2S+1}- \Phi_{i}^{ 2S+1}}
\label{calleneq}
\end{eqnarray}

where $\Phi_{i}$ is a local effective boson occupation number.  

\begin{eqnarray}
\Phi_{i}= -\frac{1}{2 \pi}\frac{1} {\langle S_{i}^{z} \rangle}\int_{-\infty}^{+\infty} \frac{ImG_{ii}(\omega)}{\exp(\omega/kT)-1} d\omega 
\end{eqnarray}

At each temperature, (5),(7) and (8) form a closed set of equations. 
These may be solved to  determine the temperature dependence of 
the local magnetization at each site and dynamical properties.

To determine the critical temperature T$_{C}$, we take the limit of vanishing 
$ \langle S_{i}^{z} \rangle \rightarrow 0$ in the previous set of equations
which leads to
\begin{eqnarray}
\langle S_{i}^{z} \rangle=\frac{1}{3}S(S+1)\frac{1}{\phi_i} 
\end{eqnarray}
We obtain,
\begin{eqnarray}
          { k_{B} T_C} &=& {1\over 3 N_{imp}} S(S+1)\sum_{i} {1\over F_{i}}
\label{eqtcrpa}
\end{eqnarray}
The local quantity $F_{i}$ is
\begin{eqnarray}
F_i \equiv \int^{+\infty}_{-\infty}{A_{ii}(E) \over {E}} dE
\end{eqnarray}
where the reduced variable $E=\frac{\omega}{m}$, m is the averaged magnetization over impurity sites,

\begin{eqnarray}
\lambda_{i} = lim_{T \rightarrow T_C} \frac{ \langle S_{i}^{z} \rangle}{m}
\end{eqnarray}

The local spectral function $A_{ii}(E)$ is,

\begin{eqnarray}
 A_{ii}(E) = -{1\over 2\pi} Im({G_{ii}(E)\over{\lambda_i}})
\end{eqnarray}

We note that by the above argument, the dependence of T$_{C}$ on quantum spin $S$ is 
entirely in the factor $S(S+1)$ in equation (\ref{eqtcrpa}). For the 
magnetization in the ordered phase the dependence is more complex
as equation (\ref{calleneq}) must be used at each stage of iteration.

In the following, to evaluate the effect of both disorder and thermal fluctuations beyond mean field theory we will compare our self-consistent local-RPA expression of the Curie temperature T$_C$ to the often-quoted MF-VCA  expression.

 \begin{eqnarray}
{\rm{T}_C}^{MF-VCA} &=& {S(S+1)\over {3}} x \sum_{i} {N_{i} J(r_{i})} 
\label{eqtcmfvca}
\end{eqnarray}
$N_{i}$ (resp. $r_{i}$) is the number (resp. distance) of the $i^{th}$ nearest neighbour (summation over all the host crystal sites). 
We note that  this expression is equivalent to equation 4 of ref\cite{DasSarma}, and  is the basis of the results of 
that paper. While in that paper sampling over different configurations
of disorder was performed, the mean-field treatment of thermal fluctuations
effectively replaced the complex geometry of 
random substitution by an homogeneous crystal lattice. 
In contrast, the SC-LRPA  simultaneously  treats the effects of random geometry explicitly and 
 includes   thermal fluctuations beyond molecular field theory.
In this case the disordered medium can no longer be reduced
to an effective crystal. In the following we will be able to see {\it quantitatively} 
how  inaccurate  such  mean-field lattice approximations  may be.

\section{Numerical results}

\subsection{Results for  Self-Consistent local RPA for the Curie Temperature
as a function of carrier density}

\begin{figure}[tbp]
\centerline{
\epsfig{file=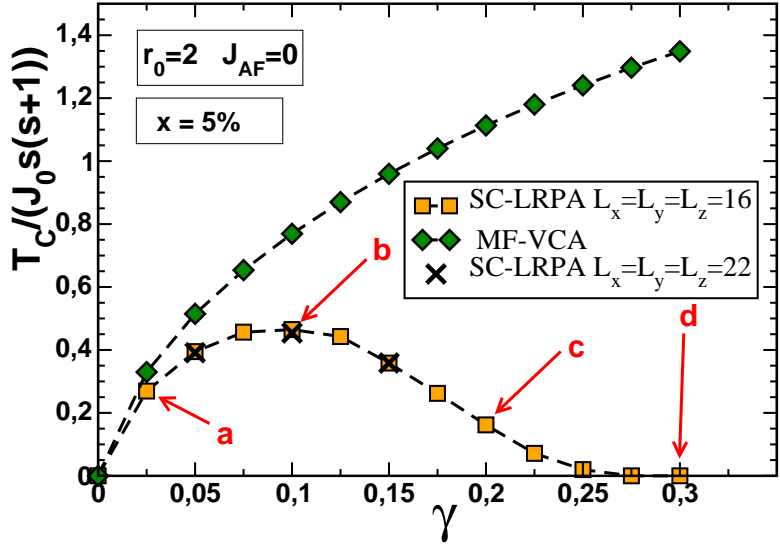,width=9cm,angle=-90}}
\caption{(Color online) Curie temperature as a function of the carrier concentration (per impurity) $\gamma$ for a fixed 5\% concentration of impurities. The damping is  $r_{0}=2$ lattice parameters and the superexchange
contribution vanishes: $J_{AF}=0$.
The Curie temperature is averaged over 100 configurations of disorder for the self-consistent local RPA.
Finite size corrections are seen to be small, by comparing  points calculated for  lattices of  $16 \times 16 \times 16$ (crosses) or  $22 \times 22 \times 22$ (squares). For comparison we show the results of MF-VCA. The points (a-d)
are indicated for reference to the following Figure~\ref{distribution}
}
\label{curiegamma}
\end{figure}

In this section we  present self-consistent  local-RPA calculations for the 
RKKY interaction for parameters, in particular concentration of interest
for III-V semiconductors. We shall compare to the results of MF-VCA treatment 
and attempt to  understand the differences when they appear.
We first consider the RKKY interaction in the absence of AF superexchange term,
i.e. we fix  $J_{AF}=0$, and assume
strong damping,  $r_{0}=2$, in units of the cubic lattice parameter.
The effects of varying these two  parameters will be considered in the following sections.
In Figure~\ref{curiegamma} we have plotted the Curie temperature for fixed concentration $x=5 \%$ of Mn impurities
as a function of $\gamma=n_{c}/x$  for both the self-consistent local-RPA and MF-VCA treatment.

\begin{figure}[tbp]
\centerline{
\epsfig{file=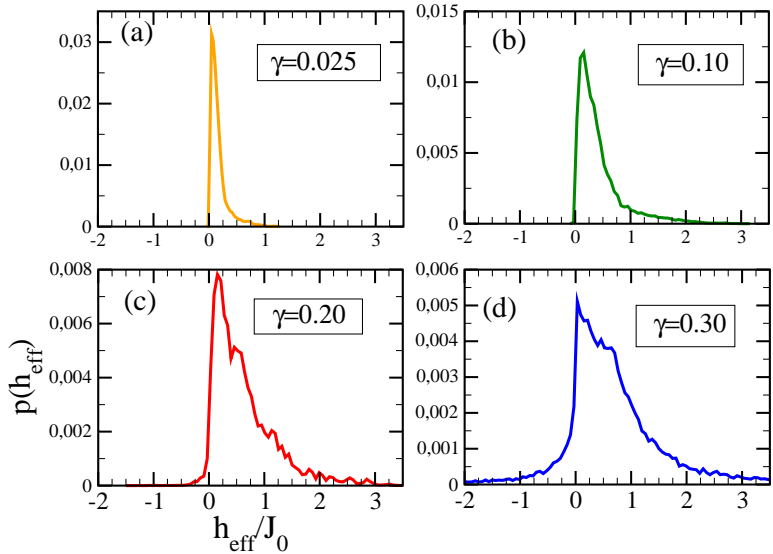,width=9cm,angle=-90}}
\caption{(Color online)  Distribution of the local fields for $\gamma= 0.025$, 0.1, 0.2 and 0.3 and the same parameters shown in Figure~\ref{curiegamma}.  The distributions correspond respectively to point (a), (b), (c) and (d) in Figure~\ref{curiegamma}.
}
\label{distribution}
\end{figure}

\begin{figure}[tbp]
\centerline{
\epsfig{file=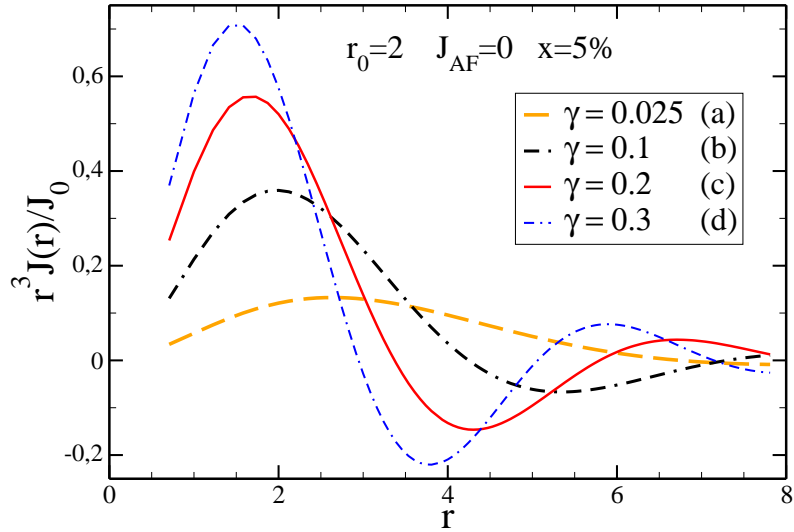,width=9cm,angle=-90}}
\caption{(Color online)  Variation of the exchange constants with   distance between
impurities for  $\gamma= 0.025$, 0.1, 0.2 and 0.3 and the same parameters shown in Figure~\ref{curiegamma}. r is in units of the lattice parameter of the fcc lattice}  
\label{exchangereal}
\end{figure}

Note that for the SC-LRPA  an average is made over approximately 100 configurations of 
disorder for each set of parameters. 
 For each configuration the magnetic impurities location have been randomly generated on a LxLxL fcc lattice, where $L \ge 16$.
In order to show that finite size effects are negligible we have plotted in the same figure the calculations  corresponding respectively to 744 and 1987 magnetic impurities distributed randomly over the fcc lattice.
By comparing results from the different sizes it is seen that finite
size effects are negligible for the  sizes considered here.
It is seen that MF-VCA systematically overestimates the Curie temperature, and
agreement between MF-VCA and SC-LRPA is only observed for very low carrier densities.
 Whilst the MF-VCA curve is monotonically increasing, the  SC-LRPA curve exhibits a maximum at relatively low density of carriers $n_{c} \approx 0.1 x$ and ferromagnetism disappears for
$\gamma \ge 0.25$.  Thus the better treatment of disorder and thermal fluctuations leads to 
a narrow region of stability for ferromagnetism. 
What is the reason for this?
In the RKKY form of the exchange integrals for sufficiently large distances the exchange 
integrals become antiferromagnetic, leading to frustration of the ferromagnetism.
The effects of this frustration are strongly underestimated within a VCA mean field treatment
This is illustrated in Fig. ~\ref{distribution} , where we plot the distribution $P(h^{eff})$ of the local fields at $T=0~K$. For a given configuration of disorder the local field at site i is
\begin{eqnarray}
h_{i}^{eff} = \sum_{l} J_{il} \langle S^{z}_{l} \rangle
\end{eqnarray}
In  Figure~\ref{distribution} the distributions are plotted for  $\gamma = 0.025 , 0.1 , 0.2, 0.3$, corresponding to (a) , (b) , (c) and (d) respectively, in Figure~\ref{curiegamma}.
In cases (a) and (b) we observe that the distribution of local fields is completely ferromagnetic: there is no  antiferromagnetic part  $P(h)=0$ for $h\le 0$.
The distribution is  very narrow in the case (a), this explains why in Figure~\ref{curiegamma} MF-VCA and SC-LRPA are very close. The distribution broadens asymmetrically  in  case (b) and the average value
$\langle h^{\rm eff} \rangle = \int_{-\infty}^{+\infty} hP(h)$ increases, leading to higher T$_C$ in MF-VCA than in SC-LRPA.
For higher density of carriers,(c) and (d), we observe a negative tail in the
local field distribution. Note that, although this tail is very small, its effect on the Curie temperature is dramatic. Figures~\ref{curiegamma} and ~\ref{distribution} show clearly that the ferromagnetism in diluted systems is very sensitive to  frustration effects coming from the exchange couplings at large distances.
In a VCA the effect of the tail is lost in the average over different sites of the lattice. In the SC-LRPA we see clearly an instability 
to bulk uniaxial ferromagnetism. 
It may seem surprising that 
a relatively small proportion of sites with local negative
molecular fields can suppress the ferromagnetism; however
by analogy to the response of a ferromagnet to random {\it external
fields} we can make an Imry-Ma argument \cite{ImryMa} that 
in three dimensions the rotational symmetry allows for 
complete break-up of ferromagnetism even for a
small proportion of random fields. Thus it seems likely that 
the resultant state is either paramagnetic or of spin-glass type.

In Figure~\ref{exchangereal} we mark the variation of the 
Heisenberg exchanges (normalized to omit the $1\over {r^3}$ decrease) as a function of distance for the same parameters  $\gamma = 0.025,  0.1, 0.2, 0.3$, corresponding to (a) , (b) , (c) and (d) respectively, in Figure~\ref{curiegamma}. Unlike the distribution of local fields, this
gives no obvious clue as to the instability of ferromagnetism.
This is perhaps surprising if we remark that the average distance
between magnetic impurities at this concentration is only slightly larger
 (r=1.07 in the units of Fig.~\ref{exchangereal})
than the lattice parameter
 of the fcc lattice.

\subsection{Effect of the exponential cut-off.}

\begin{figure}[tbp]
\centerline{
\epsfig{file=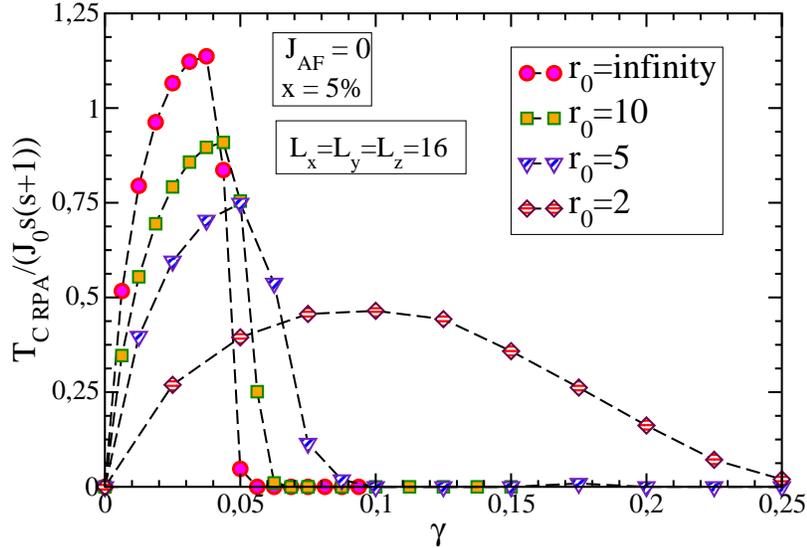,width=9cm,angle=-90}}
\caption{(Color online)  Curie temperature (self-consistent local-RPA) as a function of the carrier density 
for different values of $r_{0}$:$r_{0}$= 2; 5; 10; $\infty$. $J_{AF}=0$ and the system size is $16 \times 16 \times 16$.
}
\label{curiedamping}
\end{figure}

We now discuss the effect of the exponential cut-off on the 
domain of ferromagnetic
stability. 
The use of such a cut-off 
dates back to deGennes \cite{deGennes} who calculated the average coupling
in weakly disordered systems. There has been much discussion of whether
the difference between this average coupling and the undamped {\it typical} couplings that may
determine the characteristic scale for spin-glass behaviour
in the very dilute limit but for ferromagnetic transitions this has been 
assumed, without absolute proof to be appropriate.
In Figure~\ref{curiedamping} we have plotted the Curie temperature as a function of the carrier density 
for different values of $r_{0}$:$r_{0}$= 2; 5; 10; $\infty$. Note that  $r_{0}=\infty$ corresponds to the pure undamped RKKY case.

We  defer the question of the  influence of superexchange $J_{AF}=0$ and consider a fixed density of magnetic impurities $x=5 \%$.
The SC-LRPA shows that in the undamped RKKY limit the region of stability is 
extremely narrow. We observe that the ferromagnetic stability domain increases significantly
when $r_{0}$ is relatively small, ie of the order of the average 
distance between impurities. For instance for $r_{0}$=2, the region of stability of ferromagnetism is five times broader than in the pure undamped case. This  illustrates the strong influence of the oscillating tail on ferromagnetism.
This figure also shows clearly that the undamped RKKY exchange integrals often used in model 
calculations can not explain the ferromagnetism often observed in III-V 
diluted 
materials for the carrier densities $n_c \approx x$ or $\gamma \approx 1$ for annealed 
samples which exhibit the highest T$_{C}$. Even for  strong disorder  ( $r_{0}$=2), ferromagnetism is not possible for
 $n_c \ge 0.25 x$. From this figure and Figure~\ref{curiegamma}  , we conclude that the use of RKKY exchange integrals often used 
in simplified theories is questionable and its often-cited ``success''
is due to the unrealistic MF-VCA approximation which overestimates the 
Curie temperature and misses
the instability seen here. 
The difference with the agreement with  first principle calculations 
is that there, as in model calculations with the disorder treated
in CPA\cite{Bouzerar-unpublished}, not only are the  exchange integrals are strongly damped 
but, in addition,  the anti-ferromagnetic  contributions are almost completely suppressed\cite{Josef}.

\begin{figure}[tbp]
\centerline{
\epsfig{file=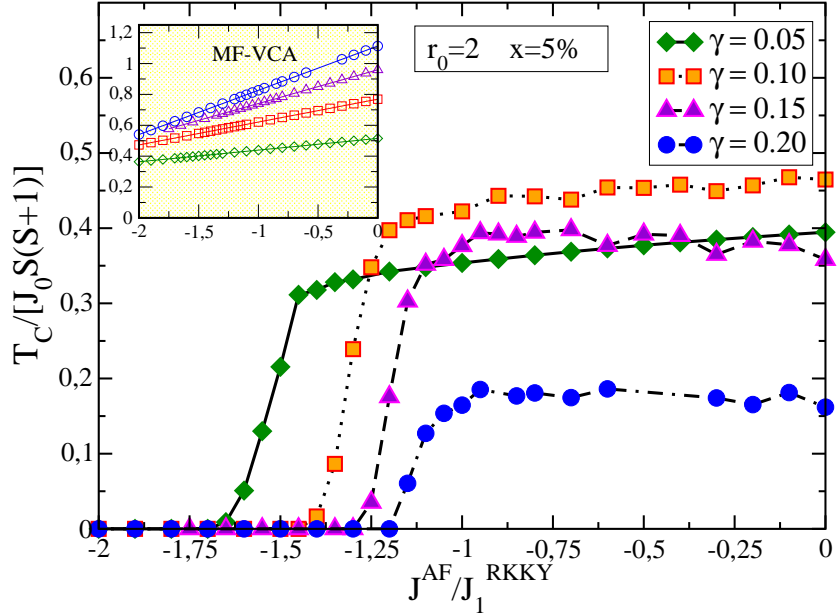,width=9cm,angle=-90}}
\caption{(Color online) SC-LRPA Curie temperature as a function of
$J_{AF}$ (normalized by the nearest-neighbour RKKY interaction $J_1$) for different carrier densities. The magnetic impurity concentration (5\%) 
and RKKY damping are fixed  $(r_{0}=2)$. In inset (note that the axes are the same) 
 we compare the 
Curie temperature calculated within MF-VCA.  
}
\label{curiesuperex}
\end{figure}
\par

\subsection{Influence of the superexchange.}

Let us now turn to  the influence on ferromagnetism of a short-range superexchange contribution. This is important for understanding materials as it is essentially
independent of the longer range RKKY exchange and sensitive to local
effects. In mean field theories a very large ferromagnetic coupling
at small distances has been used to argue, for example in doped Ga(Mn)N,
for potentially high Curie temperatures.
For simplicity, and because superexchange are short range, we consider the case
where superexchange modifies nearest-neighbour magnetic impurities only.

In Figure~\ref{curiesuperex} we have plotted the dependence of the SC-LRPA Curie temperature as a function of
$J_{AF}$ for different carrier density and fixed magnetic impurity concentration and $r_{0}$=2.
First,  in contrast to the mean-field treatment (see inset) , we observe that in the region 
dominated by the RKKY term (nearest-neighbour ferromagnetic) that the Curie temperature is {\it insensitive} to $J_{AF}$.
This can be  understood as following:  since the system is diluted, we might expect that   T$_C$ should be controlled mainly by exchange integrals for typical distances between magnetic impurities. In fact, as we have discussed,
it is not sufficient to  consider only these distances: the long-range
tail is also important.
Our local approximation builds in 
all these features. In contrast  the MF-VCA, in which the 
Curie temperature is linear in all couplings, could not
capture these geometric effects.
In our calculations, when the antiferromagnetic superexchange becomes dominant
 the Curie temperature 
vanishes abruptly. The reason for this instability  is that the system becomes frustrated, but now from the short-range
exchange, not the long range RKKY oscillations controlling the disappearance
observed in Figures \ref{curiegamma} and \ref{curiedamping}.

\subsection{Variation of T$_C$ with concentration}

\begin{figure}[tbp]
\centerline{
\epsfig{file=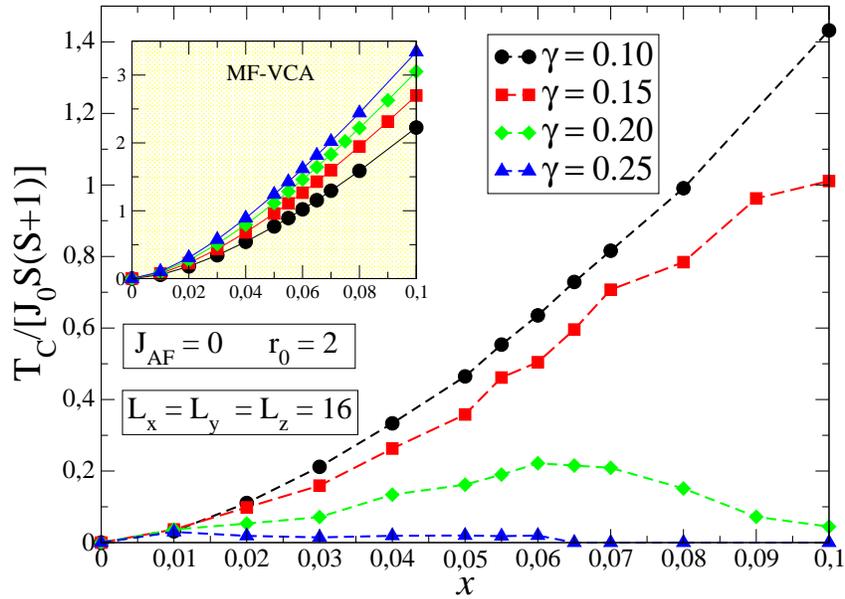,width=9cm,angle=-90}}
\caption{(Color online) SC-LRPA calculations of the Curie temperature as a function of the magnetic impurity concentration
and  for different carrier density.$J_{AF}=0$ and $r_{0}$=2. In the inset, which has the 
same axes, the MF-VCA Curie temperatures are plotted for the same parameters.
}
\label{curieconc}
\end{figure}
\par

Let us now discuss the influence of the impurity concentration on the Curie temperature.
For simplicity,
we consider only the case $J_{AF}=0$. From the previous section,  the results
change little if we add a term of short-range superexchange, provided its 
strength is insufficient to give the instability of Figure \ref{curiesuperex}.
In Figure~\ref{curieconc} we have plotted T$_{C}$ as a function of x for different values of the carrier density per impurity $n_c = \gamma x$.
For  MF-VCA calculations we observe that  T$_{C}$ increases monotonically with the density of both magnetic impurities and  itinerant carriers.
The  SC-LRPA, however, predicts much smaller  Curie temperatures than the  
MF-VCA results and that T$_C$ systematically {\it decreases} with increasing the carrier density $\gamma$.
In contrast to MF-VCA, we also observe that as we increase $\gamma$, a maximum in  T$_{C}$ appears. The location of this maximum is shifted to lower 
impurity concentration as the  carrier density per magnetic impurity  increases.
For sufficiently high densities ($\gamma \ge 0.25$) of carriers we observe  complete suppression of ferromagnetism.
We have  already discussed previously the reason for this: the  frustrating  effect of the RKKY ``tail''. 
Thus  again, we see that better treatment of disorder and fluctuations,
give very different trends from the predictions of simple effective medium
theories. The immediate question raised is then, how can we understand,
from a simple model point of view, the observed ferromagnetism? For example
that observed  at carrier
densities where from  Figure ~\ref{curieconc} we now predict $T_C=0$
and we argue that that  predictions from MF-VCA were invalid. This will be 
the subject of the next section.

\section{The solution: generalization of RKKY to treat resonances}
\begin{figure}[tbp]
\centerline{
\epsfig{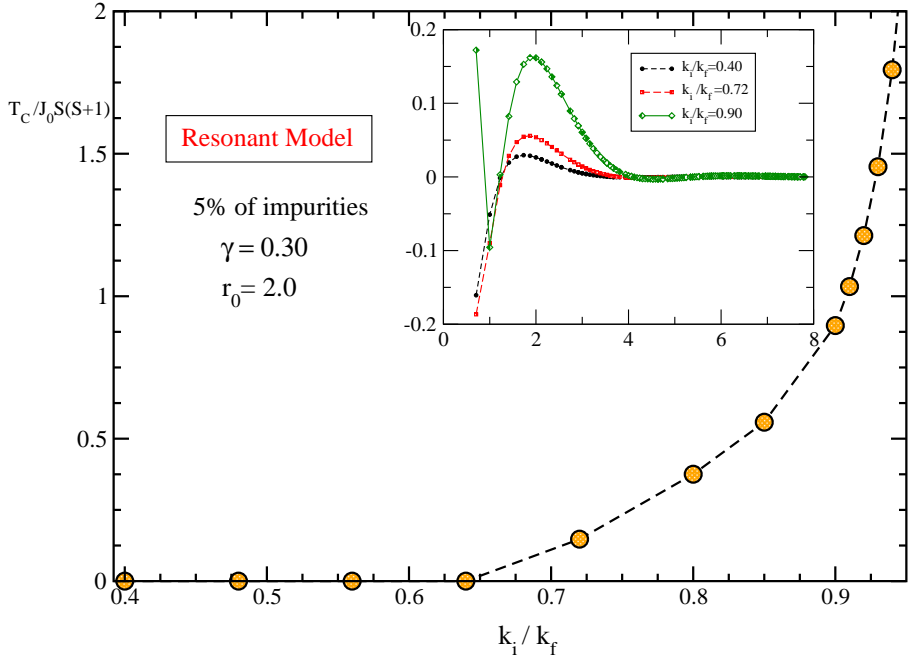}}
\caption{(Color online) Curie Temperature calculated by SC-LRPA for the model
of interacting resonances as a function of $k$-vector  ($k_i$) defining
the position of the impurity level with respect
to the Fermi vector $k_f$. Well below the Fermi level there is 
no ferromagnetism, as the resonance approaches T$_C$ becomes non-zero
and diverges as it approaches the Fermi level. In inset we show
the exchange integrals in real space for three positions of the 
resonance, to be compared to Figure ~\ref{exchangereal}.
}
\label{Fig.resonant}
\end{figure}

We have seen that simple RKKY forms of magnetic
interaction leads to a region for stable ferromagnetic order
that is very  narrow.
How can the experimental results be reconciled with this? One approach, which seems to be successful for the III-V
is to abandon a simple model approach and use {\it ab-initio} results for the effective Hamiltonian. This
works, as mentioned in the introduction, as the couplings do not display the changes of sign
characteristic of the RKKY interaction. It is interesting to look more carefully at the physical
reason behind this. 
While the non-spherical geometry of the Fermi surface
of the host material affects the oscillations in exchange coupling
it will not suppress the changes of sign.
The significant  point seems to be, in for example Ga(Mn)As, that the states at the top of the valence band are strongly
hybridized with the impurity states.
This gives rise to  resonant peaks in the density of states close to the Fermi level. 
 In consequence the states contributing to magnetic exchange
must be modified by the hybridization between impurity and host bands.
In this case the appropriate
model is that of interacting resonances\cite{Caroli,AlexanderAnderson}. 
As shown by Caroli and others \cite{Caroli,Price,Levy,ZhangLevy}, the 
simplest way 
to include  resonant effects 
in order to generalize  the regular  RKKY expression, is
to replace the term $J_{pd}$ by a frequency-dependent element ${{| V_{pd} |}^2}  \over{(\omega-\omega{+})}$. 
In perturbation theory, the interactions between two resonant impurities can be separated into  two contributions, the first an  ``RKKY-like'' term
generated by particle-hole exchange between the two spins.
While asymptotically the standard RKKY interaction
is recovered, there is an important ``pre-asymptotic'' regime where
subleading corrections to the exchange change the qualitative form 
of the interaction.
In particular in the first ``RKKY-like'' contribution there 
is  a 
ferromagnetic ``bias'', ie in the pre-asymptotic region, the interaction  oscillates with distance with period $2k_f$ around
a smoothly decaying value and  does not change sign. The second 
contribution is the  antiferromagnetic superexchange term coming
from particle-particle excitations, ie with intermediate states where the moments are unoccupied. For a perfect gas the superexchange term compensates the ferromagnetic bias of  the RKKY oscillation
giving a a total  exchange term that oscillates in sign\cite{Levy}. With the effects of disorder and interaction the compensation
need not occur\cite{Levy} and this may provide a qualitative explanation of the form of the 
exchange interactions from {\it ab-initio} approaches. Because the 
{\it ab-initio} calculations are
non-perturbative, they do  not allow an easy separation
into RKKY and superexchange contributions. As the {\it ab-initio} calculations
include interactions and, to some extent, the disorder, the similarity
between the forms seen there ( see for example Fig. 3 of Ref. \cite{Josef}) and the perturbative RKKY-like contribution to interactions
between resonances supports the idea that it makes sense to 
exclude the superexchange contribution other than the short-range
term.  
Thus an appropriate model Hamiltonian may be to take the perturbative ``RKKY-like'' interaction
between resonances. While this will be more fully
investigated in future work, we illustrate how this can restore
ferromagnetism in a range where, as  we have  noted above, standard RKKY
interactions fails to predict ferromagnetism. In Figure \ref{Fig.resonant}
we show results of calculation taking interactions which 
include the resonant RKKY-like contribution \cite{ZhangLevy,Levy}.
The resonances are defined 
by  a $k$-vector $k_i$, i.e. $E_{imp}=\hbar{\frac {k_i^2} {2m}} $
which can be varied continuously
relative to a spherical Fermi surface of fixed Fermi Energy $E_{f}=\hbar{\frac {k_f^2}{2m}}$.
Perturbative calculations between narrow resonant levels
interacting by free electrons 
give an RKKY-like contribution to the exchange\cite{ZhangLevy,Levy}, in momentum space, 
\begin{eqnarray}
 J_{RKKY}[q] = 
\frac{J_0}{4\pi q} 
\left[  
      \frac{\log | (\frac{-2\,
               {k_f} + q}{2\,{k_f} + q}
            )|}{{k_i}\,
         \left( 2\,{k_i} + q \right) } 
-\frac{
         \log | (\frac{-2\,
               {k_f} + q}{2\,{k_f} + q}
            )|}{{k_i}\,
         \left( -2\,{k_i} + q \right) } 
-      \frac{\log | (\frac{2\,
              {k_f} + q}{-2\,{k_f} + q}
           )|}{{{k_f}}^2 - {{k_i}}^2} 
\right. \nonumber \\ 
\left.    -    \frac{ \log | (\frac{{k_f} - {k_i} + q}{{k_f} - 
              {k_i}}) | }{{k_i}\,
         \left( -2\,{k_i} + q \right) } + 
      \frac{\log | (\frac{{k_f} + {k_i} + q}{{k_f} + 
              {k_i}})|}{{k_i}\,
         \left( 2\,{k_i} + q \right) } + 
      \frac{\log | (\frac{-{{k_i
                 }}^2 + 
             {\left( {k_f} + q \right) }^2}
             {{{k_f}}^2 - {{k_i}}^2})|}
         {{{k_f}}^2 - {{k_i}}^2}
       \right]
\end{eqnarray}
Interactions in real space are
obtained by Fourier transform of this analytical expression and an exponential
 damping
term to make clear comparison with the results for standard RKKY (see Figure~\ref{Fig.resonant}, inset, for the couplings as a function of distance).
We chose  the value of doping $\gamma=0.3$ and the same damping $r_0=2$, where after
Figure \ref{curiegamma} (see point ($d$) of that figure)there was no ferromagnetism for well developed
moments. It is seen that for resonances well below the Fermi level there
is no ferromagnetism, but as the resonance level increases ferromagnetism appears,
and the Curie temperature rises rapidly as the impurity
approaches the Fermi level. This is clearly a simplified
calculation, and the divergence for $k_i\rightarrow k_f$
should be suppressed if superexchange is fully included.
In addition we  assume a moment which is fixed while this
will, in fact, decrease as the resonance approaches the Fermi level.
Nevertheless it illustrates the point that it is the resonant nature of the 
exchange
that can resolve the apparent contradiction of observation, both from experiment and calculation
based on {\it ab-initio} approach, of ferromagnetism at doping levels
approaching large values of $\gamma$.
We note that in the  sense of the underlying model to take, this agrees with 
previous work emphasizing the influence of 
proximity to the Fermi level of the resonant states 
 on Curie temperature\cite{Inoue,Bouzerar1,Kikoin}.
Our calculations of the ferromagnetism are different, however,
and, we argue, more precise.

\section{Conclusions}
To conclude, we have studied the effects of both transverse fluctuations,   and disorder, 
on ferromagnetism for diluted Heisenberg models assuming  an RKKY type for the exchange integrals.
We have shown that previous MF-VCA treatments are inappropriate to study ferromagnetism in diluted magnetic systems leading to strong over-estimates of Curie
Temperatures. The apparent success of theories starting from RKKY couplings, which are often cited in the literature
as being ``qualitatively correct'', is in fact due to the oversimplified approximations (MF-VCA) used to treat the effective Heisenberg model.
We have seen from Figure \ref{curiegamma} that there is agreement only
in the limit of extremely small carrier density.
The long-range tail of the RKKY interaction destabilizes ferromagnetism
over all but very narrow ranges of parameter values. Even damping this
tail is insufficient to explain ferromagnetism at
the high doping densities of the  materials showing  the  highest Curie temperature.
 For experimentally
interesting densities, we have seen that the randomness is essential,
as seen clearly if Fig. \ref{distribution}.
Thus including the random geometry and transverse spin fluctuations, 
 as we do, is much more significant than adding  corrections to the continuum version of mean-field theory coming from 
the lattice version\cite{DasSarma}.
\par
It is appealing to have a simple phenomenological picture
of ferromagnetism
of diluted magnetic semiconductors rather than having to rely
on {\it ab-initio} calculations of exchange which must be performed
material by material and where the underlying physics may be obscured
by the fact that several aspects (hybridization, band structure, correlations, disorder...)
are  included but it is not easy to separate their effects individually.
While we have shown that  a picture of  magnetic moments
interacting with standard RKKY interactions is {\it not}
compatible with experiments, we argue that the generalization
to  include the {\it resonant} nature of the exchange may
be the correct phenomenology.


\begin{references}
\bibitem{Ohno}H. Ohno, Science {\bf 281},951 (1998).
\bibitem{Edmonds}K. W. Edmonds, P. Boguslawski,K.Y. Wang, R.P. Campion, S. N. Novikov, N.R.S. Farley,  B.L. Gallagher,
 C.T. Foxon, M. Sawicki, T. Dietl, M. Buongiorno Nardelli and J. Bernholc,  Phys. Rev. Lett. {\bf 92}, 037201 (2004).
\bibitem{Potashnik} S.J. Potashnik, K. C. Ku, S. C. Chun,J. J. Berry,N. Samarth,and P. Schiffer, Appl. Phys. Lett. {\bf 79}, 1495 (2001).
\bibitem{Dietl}  T. Dietl, H. Ohno, and F. Matsukura, Phys. Rev. B {\bf 63}, 195205 (2001).
\bibitem{DasSarma}  D.J. Priour, E.H. Hwang and S. Das Sarma, Phys Rev. Lett.  {\bf 92}, 117201 (2004).
\bibitem{Matsukura}   F. Matsukura, H.Ohno, A. Shen, and Y. Sugawara,  Phys. Rev. B {\bf 57}, R2037 (1998).
\bibitem{Sato} K. Sato, P. H. Dederichs and H. Katayama-Yoshida, Europhys Lett., {\bf 61}, 403 (2003).
\bibitem{Jungwirth}T. Jungwirth, W. A. Atkinson, B. H. Lee, and A. H. MacDonald Phys. Rev. B {\bf 59}, 9818 (1999).
\bibitem{Dogma} See for example References \cite{Dietl,Jungwirth} and references therein.
\bibitem{Wang} K.Y. Wang,K. W. Edmonds, R.P. Campion, B.L. Gallagher,N.R.S. Farley, C.T. Foxon,
M. Sawicki, P. Boguslawski, T. Dietl Journal of Applied Physics  {\bf 95}, 6512 (2004).
\bibitem{Kirby} B.J. Kirby, J.A.Borchers, , J.J. Rhyne, S.G.E. teVelthuis, A. Hoffmann, K.V. O'Donovan, T. Wojtowicz,
X. Liu, W.L. Lim and J.K. Furdyna,  Phys. Rev. B {\bf 69 }, 081307(R) (2004).
\bibitem{Sandraskii}  L.M. Sandratskii  and P. Bruno,  Phys. Rev. B {\bf 66}, 134435 (2002).
\bibitem{DDDasSarma}  P. Mahadevan, A. Zunger and D.D. Sarma, Phys Rev. Lett.  {\bf 93}, 177201 (2004).
\bibitem{Bouzerar1}  G. Bouzerar, J. Kudrnovsk\`y and P. Bruno, Phys. Rev. B {\bf 68}, 205311 (2003).
\bibitem{Bouzerar3}  G. Bouzerar, J. Kudrnovsk\`y, L. Bergqvist and P. Bruno, Phys. Rev. B {\bf 68}, 081203(R) (2003). 
\bibitem{Bouzerar-unpublished} G. Bouzerar (unpublished).
\bibitem{Brey} L. Brey and G. Gomez-Santos, Phys. Rev. B {\bf 68}, 115206 (2003).
\bibitem{Timm} C. Timm and A. H. MacDonald, Phys. Rev. B {\bf 71}, 155206 (2005).
\bibitem{Shick} A. B. Shick, J. Kudrnovsk\'y, and V. Drchal
Phys. Rev. B {\bf 69}, 125207 (2004).
\bibitem{Zhou} C. Zhou, M. P. Kennett, X. Wan, M. Berciu, R. N. Bhatt B{\bf 69} 144419 (2004).
\bibitem{Fiete}G. Fiete, G. Zarand, B. Janko, P. Redlinski, and C. P. Moca, Phys. Rev.  B{\bf 71} 115202 (2005).
\bibitem{Josef}J. Kudrnovsk\'y, I. Turek, V. Drchal, F. Maca, P. Weinberger, P. Bruno  Phys. Rev. B {\bf 69}, 115208 (2004).
\bibitem{Bouzerar2}  G. Bouzerar, T. Ziman and J. Kudrnovsk\`y, Europhys. Lett., {\bf 69}, 812-818 (2005); G. Bouzerar, T. Ziman and J. Kudrnovsk\`y, Appl. Physics Lett. {\bf 85} 4941 (2004). 
\bibitem{Bouzerar4}  Georges Bouzerar, Timothy Ziman and Josef Kudrnovsk\`y, Phys. Rev. B {\bf  72}, 125207 (2005). 
\bibitem{mc1}L. Bergqvist, O. Eriksson, J. Kudrnovsk\`y, V. Drchal, P. Korzhavyi and I. Turek, Phys. Rev. Lett., {\bf 93}
137202 (2004).
\bibitem{mc2} K. Sato, W. Schweika, P.H. Dederichs and  H. Katayama-Yoshida, Phys. Rev. B {\bf 70}, 201202(R) (2004).
\bibitem{Callen}  H.B. Callen, Phys. Rev. {\bf 130}, 890 (1963).
\bibitem{Dietl97}T. Dietl, A. Haury, and Y. Merle d'Aubign\'e, Phys. Rev. B {\bf 55}, R3347-R3350 (1997).
\bibitem{Abolfath}  M. Abolfath, T. Jungwirth, J. Brum, and A. H. MacDonald, Phys. Rev. B {\bf 63}, 054418 (2001).
\bibitem{Zarand}G. Zarand, B. Janko, Phys. Rev. Lett. {\bf 89}, 047201 (2002). 
\bibitem{ImryMa}Y. Imry and S.-K. Ma
Phys. Rev. Lett. {\bf 35}, 1399-1401 (1975).
\bibitem{deGennes}  P. G. de Gennes, J. Phys. Radium {\bf 23}, 630 (1962). 
\bibitem{Caroli}  B. Caroli, J. Phys. Chem. Solids, {\bf 28}, 1427 (1967).
\bibitem{AlexanderAnderson}  S. Alexander, P. W. Anderson, Phys. Rev. 133, A1594-A1603 (1964).
\bibitem{Price}   D. C. Price, J. Phys. F {\bf 8}, 933 (1978).
\bibitem{Levy}  Z-P Shi, P. M. Levy, J.L. Fry, Phys. Rev. B {\bf 49}, 15159 (1994).
\bibitem{ZhangLevy}  Q. Zhang, P.M.  Levy, Phys. Rev. B {\bf 34}, 1884 (1986).
\bibitem{Inoue}  J-I.  Inoue, Phys. Rev. B {\bf 67},  125302 (2003), J. Inoue, S. Nonoyama, and H. Itoh, Phys. Rev. Lett.{\bf 85}, 4610 (2000).
\bibitem{Kikoin} P. M. Krstajic, F. M. Peeters, V. A. Ivanov, V. Fleurov, and K. Kikoin, Phys. Rev. B {\bf 70}, 195215 (2004).

\end{references}
\end{document}